\documentclass[reprint,noshowpacs,noshowkeys,prd,balancelastpage,nofootinbib]{revtex4}
\usepackage{amsfonts}
\usepackage{mdframed}
\usepackage{amssymb}
\usepackage{footnote}
\usepackage{amsmath}
\usepackage{graphicx}
\usepackage{float}
\usepackage[font={footnotesize,it}]{caption}
\usepackage[utf8]{inputenc}
\usepackage{natbib}
\usepackage{tikz}
\usepackage{tikz-3dplot}
\usepackage[colorlinks=true,
            linkcolor=purple,
            urlcolor=purple,
            citecolor=blue]{hyperref}

\begin{document}

\title{Dirty black hole supported by a uniform electric field in
Einstein-Nonlinear Electrodynamics-Dilaton theory}
\author{S. Habib Mazharimousavi}
\email{habib.mazhari@emu.edu.tr}
\affiliation{Department of Physics, Faculty of Arts and Sciences, Eastern Mediterranean
University, Famagusta, North Cyprus via Mersin 10, Turkiye}

\begin{abstract}
In this study, we present an exact dirty/hairy black hole solution in the
context of gravity coupled minimally to a nonlinear electrodynamic (NED) and
a Dilaton field. The NED model is known in the literature as the square-root
(SR) model i.e., $\mathcal{L}\sim \sqrt{-\mathcal{F}}.$ The black hole
solution which is supported by a uniform radial electric field and a
singular Dilaton scalar field is non-asymptotically flat and singular with
the singularity located at its center. An appropriate transformation results
in an interesting line element $ds^{2}=-\left( 1-\frac{2M}{\rho ^{\eta ^{2}}}%
\right) \rho ^{2\left( \eta ^{2}-1\right) }d\tau ^{2}+\left( 1-\frac{2M}{%
\rho ^{\eta ^{2}}}\right) ^{-1}d\rho ^{2}+\varkappa ^{2}\rho ^{2}d\Omega
^{2} $ with two parameters - namely the mass $M$ and the Dilaton parameter $%
\eta ^{2}>1$ ($\varkappa ^{2}=\frac{1}{\eta ^{2}}$) - which may be simply
considered as the dirty Schwarzschild black hole. This is because with $\eta
^{2}\rightarrow 1$ the spacetime reduces to the Schwarzschild black hole. We
show that although the causal structure of the above spacetime is similar to
the Schwarzschild black hole, it is thermally stable for $\eta ^{2}>2$.
Furthermore, the tidal force of this black hole behaves the same as a
Schwarzschild black hole, however, its magnitude depends on $\eta ^{2}$\
such that its minimum is not corresponding to $\eta ^{2}=1$\ (Schwarzschild
limit).
\end{abstract}

\date{\today }
\maketitle

\section{Introduction}

The terminology "dirty" black hole that has been introduced by Matt Visser
in \cite{MV1} refers to black holes surrounded by some kind of classical
matter such as electromagnetic or scalar fields. In the latter case, the
black holes are also called "hairy". Therefore, in this regard the
Schwarzschild black hole which is characterized only by its mass is not
dirty, however, the Reissner-Nordstr\"{o}m black hole is a dirty one whose
dirt is the electromagnetic static field. In a system of gravity coupled to
electromagnetism, adding Dilaton \cite{RD1,RD2,RD3}, axion \cite{RA1,RA2,RA3}%
, Dilaton and axion \cite{RDA} or Abelian Higgs field \cite{RAH} results in
some interesting dirty black holes. The effects of the dirtiness matter
fields are on the physical structure of the black holes, for instance, in
the Hawking temperature \cite{MV1}, gravitational wave astronomy \cite{GW},
quasinormal modes \cite{QN1,QN2,QN3,QN4} and tidal force \cite{TF}. In \cite%
{BR1}, Bronnikov and Zaslavskii have studied generic static spherically
symmetric dirty/hairy black holes supported by an energy-momentum tensor
expressed by $T_{\mu }^{\nu }=diag\left( -\rho ,p_{r},p_{t},p_{t}\right) .$
In the latter equation, $\rho $, $p_{r},$ and $p_{t}$, respectively, are the
energy density, the radial and transverse pressure of the matter field which
is supposed to be in equilibrium with the black hole. In particular, they
investigated the equilibrium conditions between the black hole and the
classical matter field surrounding the black hole in terms of the radial
pressure to density ratio i.e., $w=\frac{p_{r}}{\rho }$ \cite{BR1}.
Accordingly, the following two cases were reported upon which the
equilibrium is possible: i) $\lim_{u\rightarrow u_{h}}w\left( u\right)
\rightarrow -1$ and ii) $\lim_{u\rightarrow u_{h}}w\left( u\right)
\rightarrow -1/\left( 1+2k\right) $ and $\rho \left( u\right) \sim \left(
u-u_{h}\right) ^{k}$, in which $u$ is the radial coordinate, $u_{h}$ is the
horizon and $k>0$. Furthermore, Bronnikov and Zaslavskii have generalized
their results for an arbitrary static spacetime in \cite{BR2} where general
static black holes in matter were considered and the case for the nonlinear
equation of state has been studied in \cite{Z1}.

Power-law Maxwell nonlinear electrodynamics (PM-NED) model was proposed in 
\cite{P1} and soon after became popular among the other models of NED \cite%
{P2,P3,P4,P5,P6,P7,P8,P9,P10,P11,P12,P13}. The model is simply given by%
\begin{equation}
\mathcal{L}=\alpha \mathcal{F}^{p}  \label{I1}
\end{equation}%
in which $\mathcal{F}=F_{\mu \nu }F^{\mu \nu }$ is the Maxwell invariant, $%
p\neq \frac{1}{2},0$ is any real number and $\alpha $ is a dimensionful
coupling constant. While the reason for excluding $p=0$ seems obvious it is
not so clear for $p=\frac{1}{2}.$ To see why let's consider Maxwell's
nonlinear equation in flat spacetime for a point electric charge sitting at
the origin. Maxwell's field of such configuration is assumed to be%
\begin{equation}
\mathbf{F}=E\left( r\right) dt\wedge dr  \label{I2}
\end{equation}%
where $E\left( r\right) $ is the radial electric field produced by the
electric charge. Maxwell's equation is given by%
\begin{equation}
d\left( \frac{\partial \mathcal{L}}{\partial \mathcal{F}}\mathbf{\tilde{F}}%
\right) =0  \label{I3}
\end{equation}%
in which%
\begin{equation}
\mathbf{\tilde{F}=}E\left( r\right) r^{2}\sin \theta d\theta \wedge d\varphi 
\label{I4}
\end{equation}%
is the dual electromagnetic field in the flat spherically symmetric
spacetime described by the line element%
\begin{equation}
ds^{2}=-dt^{2}+dr^{2}+r^{2}\left( d\theta ^{2}+\sin ^{2}\theta d\varphi
^{2}\right) .  \label{I5}
\end{equation}%
Evaluating the Maxwell invariants, one finds $\mathcal{F}=-2E^{2},$ and
consequently Eq. (\ref{I3}) implies%
\begin{equation}
r^{2}E^{2p-1}=\text{const.,}  \label{I6}
\end{equation}%
that yields%
\begin{equation}
E\left( r\right) =\frac{q}{r^{\frac{2}{2p-1}}},\text{ (}q\text{ is an
integration constant).}  \label{I7}
\end{equation}%
Clearly, (\ref{I6}) is not satisfied for $p=\frac{1}{2}$ which also causes $%
E\left( r\right) $ is undetermined for $p=\frac{1}{2}.$ Hence one has to
exclude $p=\frac{1}{2}$. We note that the same obstacle pops up when the
spacetime is curved but spherically symmetric. Apparently, $L=\alpha \sqrt{%
\mathcal{F}}$ needs special treatment and even a special name separately
from the "power low". In the literature, it is called as square-root
nonlinear electrodynamics (SR-NED) model and it was even known before the
power-law Maxwell's model. By properly adjusting the coupling constant $%
\alpha $, the electric and magnetic SR models are $L_{e}\sim \sqrt{-\mathcal{%
F}}$ and $L_{m}\sim \sqrt{\mathcal{F}}$ respectively. The magnetic model was
proposed long ago by Nielsen and Olesen \cite{NO} in string theory and was
used by 't Hooft for introducing the confinement and linear potential for
quarks \cite{GT}. Adding the electric square root term to the Maxwell linear
theory results in an electric confinement field in the black hole spacetime 
\cite{G1,G2,G3,G4,G5,G6,G7}. The SR-NED model is also the strong field limit
of the famous Born-Infeld (BI) \cite{B1,B2} model when $\mathbf{E.B}=0.$

Recently in \cite{HM1}, we have introduced a $2+1+p-$dimensional uniform
magnetic brane in the context of gravity minimally coupled with the magnetic
SR-NED. With $p=1$ the solution reduces to the Bonnor-Melvin magnetic
universe with a cosmological constant studied in \cite{MZ1}. In particular,
we have shown that the spacetime is regular and supported by a uniform
magnetic field in the sense that Maxwell's invariant is uniform.

In this research, our aim is to introduce a dirty/hairy black hole solution
in the context of Einstein's gravity coupled to SR-NED and a Dilaton scalar
field. In particular, we add the Dilaton field to come over the obstacle
that appears in the PM-NED with $p=1/2.$\ We recall that the well-known
Einstein-Maxwell-Dilaton theory admits black holes in the asymptotically
flat \cite{AF1} and non-asymptotically flat regimes \cite{YZ0}. There are
several research papers based on such a class of black holes that study the
various aspects and applications of the theory. Furthermore,
Einstein-NED-Dilaton theory with the BI-NED model has also received
attention in the literature \cite{BID,YZ1}. Considering the value of such
theories, we believe that the Einstein-SR-NED-Dilaton theory which
represents the strong field regime of the later theory will find its
applications probably in AdS/CFT correspondence. As we shall see the form of
the black hole solution is rather simple which more looks like to be a
correction in the Schwarzschild black hole. In other words, the effects of
the Dilaton and SR-NED are combined in only one additional parameter which
we shall call it its dirtiness parameter. Therefore, the final black hole
consists of two parameters in comparison with the Schwarzschild black hole
which consists of only one parameter.

Let us note that in the string theory, in the low energy limit, the problem
is described by the action%
\begin{equation}
I=\int d^{4}x\left( \mathcal{R}-\frac{1}{2}\left( \nabla \psi \right) ^{2}-%
\frac{1}{4}e^{-2\psi }F_{\mu \nu }F^{\mu \nu }\right)  \label{I8}
\end{equation}%
in which the Dilaton field is represented by $\psi $ \cite{Horowitz} (also 
\cite{Maeda}). In other words, the charged black holes in string theory are
hairy and the hair/Dilaton is coupled nonminimally to the electromagnetic
fields \cite{Green}. In the same context i.e.,
Einstein-Maxwell-Scalar/Phantom theory there have been black hole solutions
that are either asymptotically flat or non-asymptotically flat \cite%
{R1,R2,R3,R4,R5,R6,R7,R8,R9}.

The organization of the paper is as follows. In Sec. II we introduce the
theory by giving the action and the field equations. Also, we solve the
field equations exactly and present the results analytically in the same
section. In Sec. III we study the general properties of the black hole. The
physical properties consist of the energy conditions, the null geodesics,
the mass of the black hole, the thermal stability analysis, and the first
law of thermodynamics of the black hole and the causal structure. In Sec. IV
we study the tidal force of the dirty black hole and we conclude our paper
in Sec. IV.

\section{The action, the field equations and the solutions}

We start with the Einstein-Nonlinear Electrodynamic-Dilaton action described
by 
\begin{equation}
I=\int d^{4}x\left( \mathcal{R}-\frac{1}{2}\partial _{\mu }\psi \partial
^{\mu }\psi +e^{-2b\psi }\mathcal{L}\left( \mathcal{F}\right) \right) 
\label{1}
\end{equation}%
in which $b\neq 0$ is a free Dilaton parameter, $\mathcal{F}=F_{\mu \nu
}F^{\mu \nu }$ is the electromagnetic invariant with $\mathbf{F}=F_{\mu \nu
}dx^{\mu }\wedge dx^{\nu }$ the abelian electromagnetic field satisfying the
Bianchi identity 
\begin{equation}
d\mathbf{F}=0  \label{4}
\end{equation}%
and $\psi =\psi \left( r\right) $ is the Dilaton field. The nonlinear
electromagnetic Lagrangian model is given by \cite{NO,GT}%
\begin{equation}
\mathcal{L}\left( \mathcal{F}\right) =\alpha \sqrt{-\mathcal{F}}  \label{L}
\end{equation}%
where $\alpha $ is a dimensionful constant parameter. Variation of the
action with respect to the metric tensor gives Einstein's field equations
expressed by%
\begin{equation}
\mathcal{R}_{\mu }^{\nu }=2\partial _{\mu }\psi \partial ^{\nu }\psi +\frac{%
\alpha e^{-2b\psi }}{\sqrt{-\mathcal{F}}}F_{\mu \lambda }F^{\nu \lambda }.
\label{E}
\end{equation}%
Furthermore, variation of the action with respect to the Dilaton scalar
field yields the Dilaton field equation given by 
\begin{equation}
\nabla _{\mu }\nabla ^{\mu }\psi \left( r\right) =\frac{\alpha be^{-2b\psi }%
}{2}\sqrt{-\mathcal{F}}  \label{2}
\end{equation}%
and finally, variation of the action with respect to the gauge potential
yields the NED-Dilaton equation%
\begin{equation}
d\left( \frac{e^{-2b\psi }}{\sqrt{-\mathcal{F}}}\mathbf{\tilde{F}}\right) =0
\label{3}
\end{equation}%
in which $\mathbf{\tilde{F}}$ is the dual field two-form of $\mathbf{F}.$
Our spacetime is static and spherically symmetric with the line element%
\begin{equation}
ds^{2}=-f\left( r\right) dt^{2}+\frac{dr^{2}}{f\left( r\right) }+R\left(
r\right) ^{2}d\Omega ^{2},  \label{5}
\end{equation}%
where $d\Omega ^{2}=d\theta ^{2}+\sin ^{2}\theta d\varphi ^{2}$ is the line
element on the 2-sphere. The electromagnetic field is chosen to be a pure
electric field produced by a point charge sitting at the origin expressed by%
\begin{equation}
\mathbf{F}=E\left( r\right) dt\wedge dr  \label{6}
\end{equation}%
with its dual field obtained to be%
\begin{equation}
\mathbf{\tilde{F}}=E\left( r\right) R\left( r\right) ^{2}\sin \theta d\theta
\wedge d\varphi .  \label{7}
\end{equation}%
The electrodynamic invariant is obtained to be%
\begin{equation}
\mathcal{F}=-2E\left( r\right) ^{2}  \label{Inv}
\end{equation}%
upon which the NED-Dilaton equation (\ref{3}) implies%
\begin{equation}
R\left( r\right) ^{2}=R_{0}^{2}e^{2b\psi }.  \label{8}
\end{equation}%
where $R_{0}$ is an integration constant related to the electric charge of
the electric monopole. We add that Eq. (\ref{8}) has been found through the
NED-Dilaton equation (\ref{3}) explicitly. In a similar situation with
linear or nonlinear electrodynamics in the literature, such a relation is
usually considered in the form of an ansatz. For instance, we refer to \cite%
{RDEH1,RDEH2}. Another important observation regarding the NED-Dilaton
equation (\ref{3}) is that it doesn't identify the form of the electric
field $E(r)$\ and on the contrary $E(r)$\ canceled out. This, however,
doesn't mean that $E(r)$\ is an arbitrary function. As we shall see it will
be identified from the other field equations, uniquely. \ 

Einstein's field equations are given by 
\begin{equation}
\mathcal{R}_{t}^{t}=-\frac{\alpha R_{0}^{2}E\left( r\right) }{\sqrt{2}%
R\left( r\right) ^{2}},  \label{E0}
\end{equation}%
\begin{equation}
\mathcal{R}_{r}^{r}=2f\left( \psi ^{\prime }\right) ^{2}+\mathcal{R}_{t}^{t},
\label{E1}
\end{equation}%
and%
\begin{equation}
\mathcal{R}_{\theta }^{\theta }=\mathcal{R}_{\varphi }^{\varphi }=0,
\label{E2}
\end{equation}%
where%
\begin{equation}
\mathcal{R}_{t}^{t}=-\frac{1}{2R}\left( Rf^{\prime \prime }+2f^{\prime
}R^{\prime }\right) ,  \label{R00}
\end{equation}%
\begin{equation}
\mathcal{R}_{r}^{r}=-\frac{1}{2R}\left( Rf^{\prime \prime }+4fR^{\prime
\prime }+2f^{\prime }R^{\prime }\right)  \label{R11}
\end{equation}%
and%
\begin{equation}
\mathcal{R}_{\theta }^{\theta }=\mathcal{R}_{\varphi }^{\varphi }=-\frac{1}{%
R^{2}}\left( RR^{\prime }f^{\prime }+fRR^{\prime \prime }-1+\left( R^{\prime
}\right) ^{2}f\right)  \label{R22}
\end{equation}%
are Ricci tensor's components. Furthermore, the Dilaton field equation (\ref%
{2}) explicitly becomes%
\begin{equation}
\left( R^{2}f\psi ^{\prime }\right) ^{\prime }=\frac{\alpha bR_{0}^{2}\sqrt{2%
}E\left( r\right) }{2}.  \label{9}
\end{equation}%
Note that wherever needed we used (\ref{8}) to simplify the equations. Next,
we solve (\ref{E2}) for $f\left( r\right) $ which is given by%
\begin{equation}
f\left( r\right) =\frac{r-r_{0}}{RR^{\prime }}  \label{metric}
\end{equation}%
in which $r_{0}$ is an integration constant. We are, then, left with three
equations, namely, (\ref{E0}), (\ref{E1}) and (\ref{9}), and three unknown
functions i.e., $R\left( r\right) ,$ $E\left( r\right) $ and $\psi \left(
r\right) .$ Eq. (\ref{E1}) simply becomes%
\begin{equation}
R^{\prime \prime }+R\psi ^{\prime 2}=0
\end{equation}%
which implies%
\begin{equation}
\psi ^{\prime }=\pm \sqrt{-\frac{R^{\prime \prime }}{R}},  \label{S}
\end{equation}%
in which without loss of generality we continue with the positive $\psi
^{\prime }.$ Furthermore, from Eq. (\ref{E0}) one obtains 
\begin{equation}
E\left( r\right) =-\frac{\sqrt{2}R^{2}}{\alpha R_{0}^{2}}R_{t}^{t}
\label{El}
\end{equation}%
which together with (\ref{S}) simplify Eq. (\ref{9}) into%
\begin{equation}
\left( 2b\sqrt{-R^{\prime \prime }R}-2R^{\prime }\right) \left( RR^{\prime
}R^{\prime \prime \prime }\left( r-r_{0}\right) -2R^{\prime \prime }\left(
RR^{\prime \prime }\left( r-r_{0}\right) -\frac{1}{2}R^{\prime }\left(
\left( r-r_{0}\right) R^{\prime }+2R\right) \right) \right) =0.
\label{MASTER}
\end{equation}%
The latter yields the following two possibilities:%
\begin{equation}
2b\sqrt{-R^{\prime \prime }R}-2R^{\prime }=0  \label{First}
\end{equation}%
or%
\begin{equation}
RR^{\prime }R^{\prime \prime \prime }\left( r-r_{0}\right) -2R^{\prime
\prime }\left( RR^{\prime \prime }\left( r-r_{0}\right) -\frac{1}{2}%
R^{\prime }\left( \left( r-r_{0}\right) R^{\prime }+2R\right) \right) =0.
\label{Sec}
\end{equation}%
Considering the first equation i.e., (\ref{First}) one finds%
\begin{equation}
R\left( r\right) =(C_{1}r+C_{2})^{\frac{b^{2}}{b^{2}+1}}  \label{R}
\end{equation}%
in which $C_{1}$ and $C_{2}$ are two integration constants. To keep the
solution physical one has to assume $C_{1}>0,C_{2}\geq 0$ such that $%
-R^{\prime \prime }R$ remains positive. The latter equation further implies%
\begin{equation}
E\left( r\right) =\frac{\sqrt{2}}{\alpha b^{2}R_{0}^{2}}  \label{EF}
\end{equation}%
which is uniform and%
\begin{equation}
\psi =\psi _{0}+\frac{b}{b^{2}+1}\ln \left( C_{1}r+C_{2}\right)  \label{psi}
\end{equation}%
where $\psi _{0}$ is an integration constant. Finally, the metric function $%
f\left( r\right) $ is obtained from (\ref{metric}) to be%
\begin{equation}
f\left( r\right) =\frac{\left( b^{2}+1\right) \left( r-r_{0}\right) }{%
b^{2}C_{1}}\left( C_{1}r+C_{2}\right) ^{\frac{1-b^{2}}{1+b^{2}}}.
\end{equation}%
Here we note that not only the radial electric field is constant/uniform but
Maxwell's invariant of the theory i.e., $F$\ is also uniform and given by%
\begin{equation}
\mathcal{F}=-2\left( \frac{\sqrt{2}}{\alpha b^{2}R_{0}^{2}}\right) ^{2}.
\end{equation}%
The constant $R_{0}$\ can be identified through the solution (\ref{psi}) and
its consistency with (\ref{8}), which implies 
\begin{equation}
R_{0}=e^{-b\psi _{0}}.  \label{CON}
\end{equation}%
As we can see, $R_{0}$\ doesn't appear in the rest of the field equations
and the solutions directly and its existence is through $\psi _{0}.$\
Concerning (\ref{psi}) shifting $\psi $\ by $-\psi _{0}$\ doesn't change the
kinematic term in the action (\ref{1}). Moreover, in the term regarding the
coupling of the NED and the Dilaton field i.e.,%
\begin{equation}
e^{-2b\psi }\mathcal{L}\left( \mathcal{F}\right) =\frac{2}{b^{2}}\left(
C_{1}r+C_{2}\right) ^{\frac{-b^{2}}{b^{2}+1}},
\end{equation}%
the effects of $-\psi _{0}$\ and $R_{0}^{2}$\ have also been canceled out,
mutually. These all suggest that we set $\psi _{0}=0$\ which results in $%
R_{0}=1.$

Finally, the other possibility given by Eq. (\ref{Sec}) admits an exact
solution in the form%
\begin{equation}
R\left( r\right) =C_{3}\sqrt{\frac{\left( 1+C_{1}^{2}\right) \left(
r_{0}^{2}+C_{2}^{2}\right) }{4}+\frac{\left( C_{1}^{2}-1\right) r_{0}C_{2}}{2%
}+r^{2}+r\left( C_{2}-r_{0}\right) }\exp \left( -\frac{1}{C_{1}}\arctan
\left( \frac{2r+C_{2}-r_{0}}{C_{1}\left( C_{2}+r_{0}\right) }\right) \right)
,  \label{SS}
\end{equation}%
where $C_{1},$ $C_{2}$ and $C_{3}$ are some integration constant. Using (\ref%
{SS}) one obtains $R^{\prime \prime }>0$ and consequently from (\ref{S}) $%
\psi ^{\prime }\left( r\right) ^{2}<0$ which implies that the scalar field
is actually a phantom field. Hence we exclude this solution at least in this
current study.

\section{Physical properties of the solution}

The exact solutions of the field equations can be summarized as follows. The
electric field is radial with uniform magnitude given by Eq. (\ref{EF}) and
the spacetime is found to be described by the line element%
\begin{equation}
ds^{2}=-\frac{\left( b^{2}+1\right) \left( r-r_{0}\right) }{b^{2}C_{1}}%
\left( C_{1}r+C_{2}\right) ^{\frac{1-b^{2}}{1+b^{2}}}dt^{2}+\frac{dr^{2}}{%
\frac{\left( b^{2}+1\right) \left( r-r_{0}\right) }{b^{2}C_{1}}\left(
C_{1}r+C_{2}\right) ^{\frac{1-b^{2}}{1+b^{2}}}}+(C_{1}r+C_{2})^{\frac{2b^{2}%
}{b^{2}+1}}d\Omega ^{2}.  \label{LE2}
\end{equation}%
To see the structure of this spacetime, we apply the following
transformation $(C_{1}r+C_{2})^{\frac{b^{2}}{b^{2}+4}}\rightarrow R$ upon
which (\ref{LE2}) becomes%
\begin{equation}
ds^{2}=-\frac{\left( b^{2}+1\right) R^{\frac{1-b^{2}}{b^{2}}}}{b^{2}C_{1}}%
\left[ \frac{1}{C_{1}}\left( R^{\frac{b^{2}+1}{b^{2}}}-C_{2}\right) -r_{0}%
\right] dt^{2}+\frac{\left( \frac{b^{2}+1}{b^{2}C_{1}}\right) R^{\frac{%
b^{2}+1}{b^{2}}}dR^{2}}{\left[ \frac{1}{C_{1}}\left( R^{\frac{b^{2}+1}{b^{2}}%
}-C_{2}\right) -r_{0}\right] }+R^{2}d\Omega ^{2}.  \label{LE3}
\end{equation}%
This is easily seen that within redefinition of $r_{0}\rightarrow -\frac{%
C_{2}}{C_{1}}+\frac{R_{+}^{\frac{b^{2}+1}{b^{2}}}}{C_{1}}$ and $t\rightarrow
C_{1}t$ one can eliminate $C_{2}$ such that (\ref{LE3}) becomes%
\begin{equation}
ds^{2}=-\eta ^{2}\left( 1-\left( \frac{R_{+}}{R}\right) ^{\eta ^{2}}\right)
R^{\frac{2}{b^{2}}}dt^{2}+\frac{\eta ^{2}dR^{2}}{1-\left( \frac{R_{+}}{R}%
\right) ^{\eta ^{2}}}+R^{2}d\Omega ^{2}  \label{M3}
\end{equation}%
in which $R_{+}$ is the new constant in place of $r_{0},$ and $\eta ^{2}=%
\frac{b^{2}+1}{b^{2}}$. The spacetime described by Eq. (\ref{M3}) is a black
hole whose event horizon is located at $R=R_{+}.$ On the other hand with the
transformation $(C_{1}r+C_{2})^{\frac{b^{2}}{b^{2}+1}}\rightarrow R$ the
scalar field simplifies as%
\begin{equation}
\psi =\frac{1}{b}\ln R.  \label{D1}
\end{equation}%
We want to emphasize that $b=0$ has already been excluded which removes our
worries in (\ref{D1}). As a matter of fact, with $b=0$ the nonlinear Maxwell
equation (\ref{3}) implies $R=R_{0}$ which doesn't satisfy Einstein's
equations.

\subsection{Energy conditions}

One of the physical constraints on any matter field supporting a black hole
is the satisfying of the energy conditions. These energy conditions imply
whether the matter fields are normal or exotic. Let us write Einstein's
equation in the following form ($8\pi G=1$)%
\begin{equation}
G_{\mu }^{\nu }=T_{\mu }^{\nu }  \label{EE}
\end{equation}%
in which the effective energy-momentum tensor is written as%
\begin{equation}
T_{\mu }^{\nu }=diag\left[ -\rho ,P_{r},P_{t},P_{t}\right]  \label{EMT}
\end{equation}%
where $\rho ,$ $P_{r},$ and $P_{t}$ are the effective energy density,
radial, and transverse pressure densities, respectively. Applying Einstein
equation (\ref{EE}) and getting help from Eq. (\ref{E}) with the line
element given by Eq. (\ref{M3}), one obtains%
\begin{equation}
\rho =\frac{\eta ^{2}-1}{\eta ^{2}R^{2}}\left( 1-\left( \frac{R_{+}}{R}%
\right) ^{\eta ^{2}}\right) ,  \label{EM1}
\end{equation}%
\begin{equation}
P_{r}=\frac{\eta ^{2}-1}{\eta ^{2}R^{2}}\left( 1-\left( \frac{R_{+}}{R}%
\right) ^{\eta ^{2}}\right) ,  \label{EM2}
\end{equation}%
and%
\begin{equation}
P_{t}=\frac{\left( \eta ^{2}-1\right) ^{2}}{\eta ^{2}R^{2}}\left( 1-\frac{1}{%
\eta ^{2}-1}\left( \frac{R_{+}}{R}\right) ^{\eta ^{2}}\right) .  \label{EM3}
\end{equation}%
Thechnically, having the RHS of the Eqs. (\ref{EM1}) and (\ref{EM2}) the
same is due to the solution of the field equations which is reflected in the
line element (\ref{M3}). Recalling $\eta ^{2}>1,$ outside the black hole
where the energy-momentum tensor is described by (\ref{EMT}), all components
i.e., $\rho ,$ $P_{r},$ and $P_{t}$ are positive definite. Therefore, the
null-energy condition (NEC) implying $\rho +P_{i}\geq 0$, the weak-energy
condition (WEC) implying $\rho \geq 0$ and $\rho +P_{i}\geq 0,$ and the
strong-energy condition (SEC) implying $\rho +\sum P_{i}\geq 0$ are all
satisfied. Furthermore, the effective energy-momentum tensor vanishes at the
horizon indicating the regularity of the horizon \cite{BR1}.

In terms of the results obtained in \cite{BR1}, we would like to add that $w=%
\frac{\rho }{P_{r}}=1$ which indicates the matter field is normal, however,
both $\rho $ and $p_{r}$ become zero at the horizon.

\begin{figure}[tbp]
\centering\includegraphics[width=0.5\textwidth]{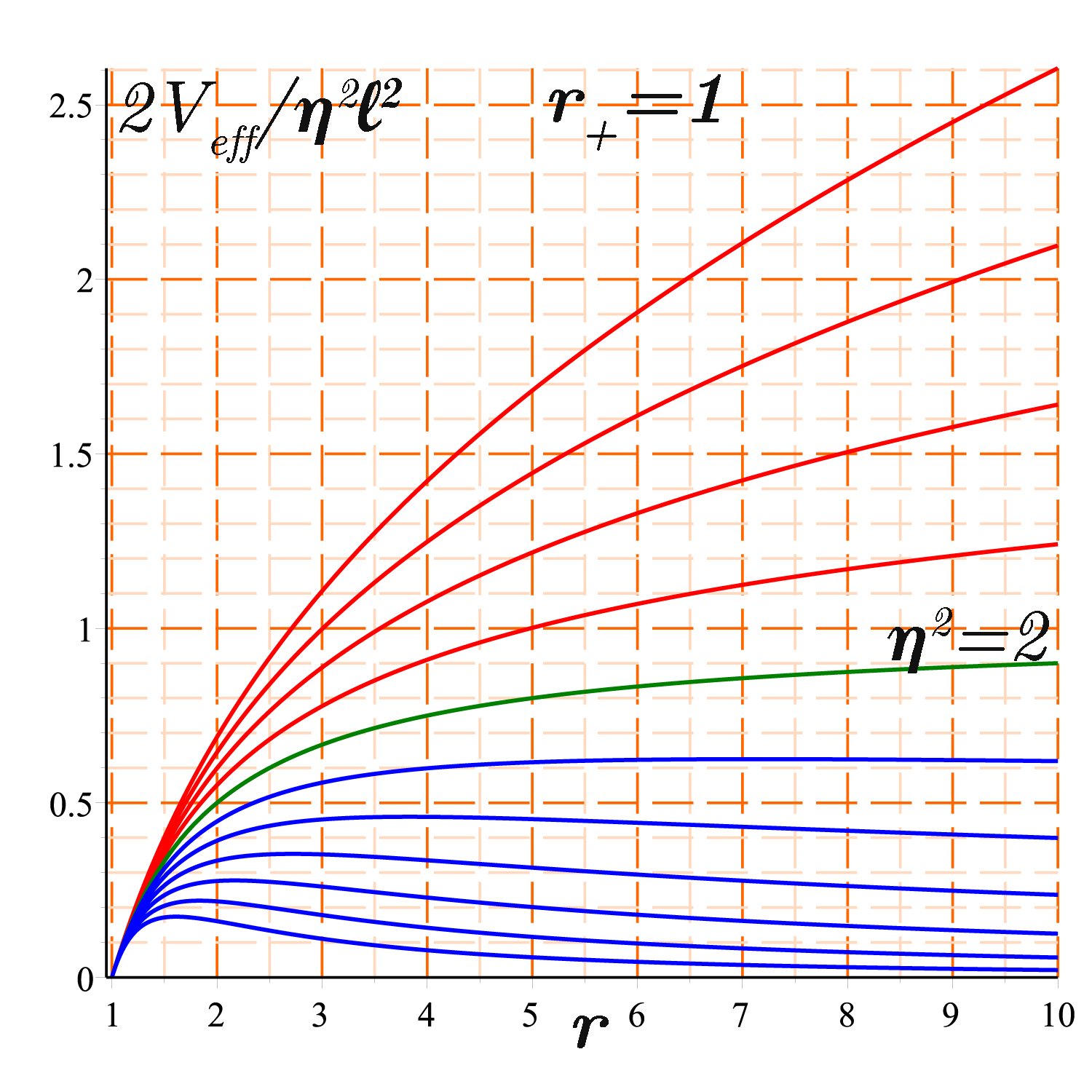}
\caption{The plots of the effective potential (\protect\ref{EP}) versus $r$
for $r_{+}=1$ and $\protect\eta ^{2}=1.1,...,2.6$ with equal steps from
bottom to top. The curve corresponding to $\protect\eta ^{2}=2.0$ is
indicated. }
\label{Fig1}
\end{figure}

\begin{figure}[tbp]
\centering\includegraphics[width=0.5\textwidth]{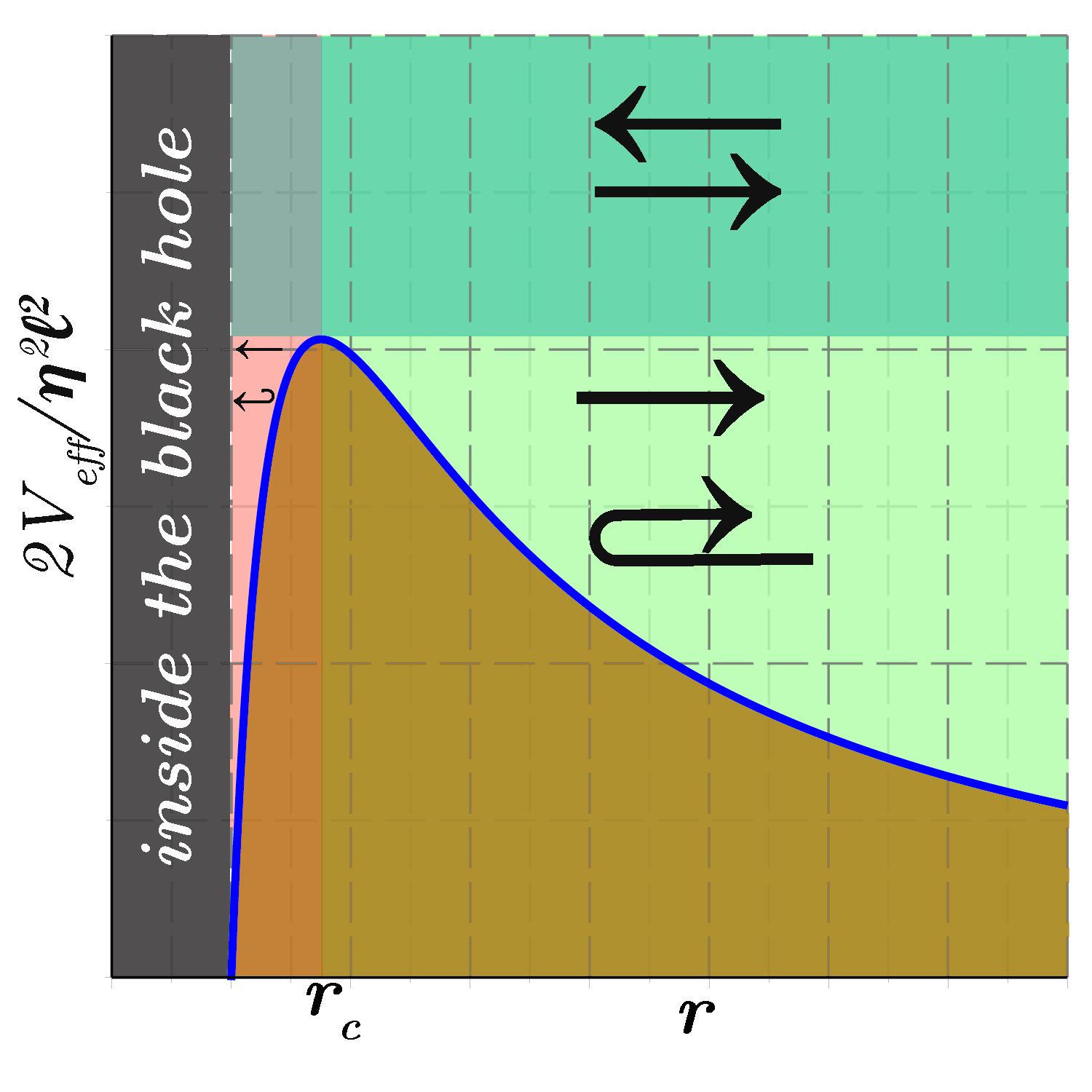}
\caption{A typical plot of the effective potential (\protect\ref{EP}) versus 
$r$ when $\protect\eta ^{2}<2.$ The potential admits an absulot maximum at $%
r=r_{c}$ where $V_{eff}\left( r_{c}\right) =\frac{\protect\eta ^{4}\ell ^{2}%
}{2\left( 4-\protect\eta ^{2}\right) }\left( \frac{4-\protect\eta ^{2}}{%
2\left( 2-\protect\eta ^{2}\right) }r_{+}\right) ^{\frac{2\left( \protect%
\eta ^{2}-2\right) }{\protect\eta ^{2}}}.$ The fate of a null particle
moving in the vicinity of this potential depends strongly on its initial
conditions and the conserved quantities. All cases are stated with arrows in
different regions that are highlighted with different colors.}
\label{Fig2}
\end{figure}

\subsection{Null geodesic}

In this section, we would like to study the photons' motion in the vicinity
of the obtained black hole (\ref{M3}). The Lagrangian of a null-particle
moving in the vicinity of the black hole (\ref{M3}) is given by%
\begin{equation}
L=-\frac{1}{2}\eta ^{2}\left( 1-\left( \frac{R_{+}}{R}\right) ^{\eta
^{2}}\right) R^{\frac{2}{b^{2}}}\dot{t}^{2}+\frac{1}{2}\frac{\eta ^{2}R^{2}}{%
1-\left( \frac{R_{+}}{R}\right) ^{\eta ^{2}}}+\frac{1}{2}R^{2}\left( \dot{%
\theta}^{2}+\sin ^{2}\theta \dot{\varphi}^{2}\right)
\end{equation}%
in which a dot stands for the derivative with respect to an affine
parameter. Considering the conserved energy 
\begin{equation}
\mathcal{E}=-\frac{\partial L}{\partial \dot{t}}=\eta ^{2}\left( 1-\left( 
\frac{R_{+}}{R}\right) ^{\eta ^{2}}\right) R^{\frac{2}{b^{2}}}\dot{t}
\end{equation}%
and the angular momentum%
\begin{equation}
\ell =\frac{\partial L}{\partial \dot{\varphi}}=R^{2}\sin ^{2}\theta \dot{%
\varphi}
\end{equation}%
together with the null condition i.e., 
\begin{equation}
g^{\mu \nu }\dot{x}_{\mu }\dot{x}_{\nu }=0
\end{equation}%
one obtains the main geodesic equation given by%
\begin{equation}
R^{2\left( \eta ^{2}-1\right) }R^{2}+\frac{\ell ^{2}}{\eta ^{2}}\left(
1-\left( \frac{R_{+}}{R}\right) ^{\eta ^{2}}\right) R^{2\left( \eta
^{2}-2\right) }=\frac{\mathcal{E}^{2}}{\eta ^{4}},  \label{GE}
\end{equation}%
where we have assumed $\theta =\frac{\pi }{2}.$ Introducing $r=R^{\eta ^{2}}$
the geodesic equation (\ref{GE}) simplifies significantly as expressed by%
\begin{equation}
\dot{r}^{2}+\eta ^{2}\ell ^{2}\left( 1-\frac{r_{+}}{r}\right) r^{\frac{%
2\left( \eta ^{2}-2\right) }{\eta ^{2}}}=\mathcal{E}^{2}.
\end{equation}%
This is in analogy with the equation of motion of a unit-mass
one-dimensional particle with mechanical energy $\frac{1}{2}E^{2}$ and
effective one-dimensional potential 
\begin{equation}
V_{eff}\left( r\right) =\frac{1}{2}\eta ^{2}\ell ^{2}\left( 1-\frac{r_{+}}{r}%
\right) r^{\frac{2\left( \eta ^{2}-2\right) }{\eta ^{2}}}.  \label{EP}
\end{equation}%
In Fig. \ref{Fig1} we plot $\frac{V_{eff}\left( r\right) }{\frac{1}{2}\eta
^{2}\ell ^{2}}$ in terms of $r$ for $r_{+}=1$ and various values of $\eta
^{2}.$ This figure displays that for $\eta ^{2}\geq 2$ irrespective of the
value of $E^{2}$ and $\ell ^{2}$ the photon falls into the singularity. On
the other hand for $\eta ^{2}<2$ the effective potential admits a maximum at 
$r=r_{c}$ where%
\begin{equation}
r_{c}=\frac{4-\eta ^{2}}{2\left( 2-\eta ^{2}\right) }r_{+}
\end{equation}%
and 
\begin{equation}
V_{eff}\left( r_{c}\right) =\frac{\eta ^{4}\ell ^{2}}{2\left( 4-\eta
^{2}\right) }\left( \frac{4-\eta ^{2}}{2\left( 2-\eta ^{2}\right) }%
r_{+}\right) ^{\frac{2\left( \eta ^{2}-2\right) }{\eta ^{2}}}.
\end{equation}%
As it is depicted in Fig. \ref{Fig2}, with $\frac{\mathcal{E}^{2}}{2\ell ^{2}%
}<V_{eff}\left( r_{c}\right) $ and $r_{0}<r_{c\text{ }}$ the photon falls
into the singularity. Furthermore, with $\frac{\mathcal{E}^{2}}{2\ell ^{2}}%
<V_{eff}\left( r_{c}\right) $ and $r_{0}>r_{c\text{ }}$the photon definitely
bounces back to infinity. Finally, for $\frac{\mathcal{E}^{2}}{2\ell ^{2}}%
>V_{eff}\left( r_{c}\right) $ the null particle either falls to the
singularity or escapes to infinity depending on its initial condition.

\subsection{The mass of the black hole}

The line element (\ref{M3}) represents a singular non-asymptotically flat
black hole. Being non-asymptotically flat implies that the standard ADM mass
is not defined for such a black hole. Hence we follow the Brown and York
(BY) (\cite{BY1,YZ0,YZ1}) formalism to introduce the so-called "quasilocal
(QL) mass". According to BY formalism, for a non-asymptotically flat line
element 
\begin{equation}
ds^{2}=-F\left( R\right) ^{2}dt^{2}+\frac{dR^{2}}{G\left( R\right) ^{2}}%
+R^{2}d\Omega ^{2}
\end{equation}%
the QL mass is given by%
\begin{equation}
M_{QL}=\lim_{R_{B}\rightarrow \infty }R_{B}F\left( R_{B}\right) \left[
G_{ref}\left( R_{B}\right) -G\left( R_{B}\right) \right]  \label{QLM}
\end{equation}%
in which $G_{ref}\left( R_{B}\right) $ is an arbitrary non-negative
reference function, which yields the zero of the energy for the background
spacetime, and $R_{B}$ is the radius of the space-like hypersurface. For the
line element (\ref{M3}) one finds $F\left( R\right) ^{2}=\eta ^{2}\left(
1-\left( \frac{R_{+}}{R}\right) ^{\eta ^{2}}\right) R^{\frac{2}{b^{2}}},$ $%
G\left( R\right) ^{2}=\frac{1-\left( \frac{R_{+}}{R}\right) ^{\eta ^{2}}}{%
\eta ^{2}}$ and $G_{ref}\left( R_{B}\right) ^{2}=\frac{1}{\eta ^{2}}$ which
result in 
\begin{equation}
M_{QL}=\frac{1}{2}R_{+}^{\eta ^{2}}.
\end{equation}%
The line element therefore becomes%
\begin{equation}
ds^{2}=-\eta ^{2}\left( 1-\frac{2M_{QL}}{R^{\eta ^{2}}}\right) R^{\frac{2}{%
b^{2}}}dt^{2}+\frac{\eta ^{2}dR^{2}}{1-\frac{2M_{QL}}{R^{\eta ^{2}}}}%
+R^{2}d\Omega ^{2}.  \label{LD2}
\end{equation}%
The latter line element gives the correct spacetime limit as $b\rightarrow
\infty $ such that $\eta ^{2}\rightarrow 1$, $E\left( r\right) \rightarrow 0$
and the solution becomes the standard Schwarzschild black hole and $M_{QL}$
turns to be identified as the ADM mass of the black hole. Moreover, by
scaling the time $t$ and the radial coordinate $R$ the latter line element
becomes%
\begin{equation}
ds^{2}=-\left( 1-\frac{2M}{\rho ^{\eta ^{2}}}\right) \rho ^{2\left( \eta
^{2}-1\right) }d\tau ^{2}+\left( 1-\frac{2M}{\rho ^{\eta ^{2}}}\right)
^{-1}d\rho ^{2}+\varkappa ^{2}\rho ^{2}d\Omega ^{2}  \label{LE}
\end{equation}%
in which $\rho =\eta R,$ $M=M_{QL}\eta ^{\eta ^{2}},$ $\tau =\eta ^{1-\frac{2%
}{b^{2}}}t$ and $\varkappa ^{2}=\frac{1}{\eta ^{2}}$. Having known that $%
\eta ^{2}=\frac{b^{2}+1}{b^{2}}>1$, (\ref{LE}) clearly implies that the
Dilaton field causes a sort of conical structure with the deficit angle
represented by $\varkappa ^{2}<1.$

\subsection{Thermal stability and the first law}

To investigate the thermal stability of the black hole (\ref{LD2}), we
calculate the Hawking temperature defined by%
\begin{equation}
T_{H}=\left( -\frac{g_{tt}^{\prime }}{4\pi \sqrt{-g_{tt}g_{rr}}}\right)
_{R=R_{+}}=\frac{\eta ^{2}}{4\pi }R_{+}^{\eta ^{2}-2}  \label{R18}
\end{equation}%
and upon applying the so-called area law \cite{A1}, the entropy of the black
hole is also given by%
\begin{equation}
S=\pi R_{+}^{2}.  \label{R19}
\end{equation}%
Finally, we calculate the specific heat capacity defined by%
\begin{equation}
C=T_{H}\frac{\partial S}{\partial T_{H}}=\frac{2\pi }{\eta ^{2}-2}R_{+}^{2}.
\label{R20}
\end{equation}%
The black hole is considered to be thermally stable if both $T_{H}$ and $C$
are positive. Therefore with $\eta ^{2}-2>0$ ($b^{2}<1$) the black hole is
thermally stable. In accordances with (\ref{R18}), with $\eta ^{2}=2$ the
Hawking temperature becomes constant which justifies the infinite heat
capacity.

Finally, knowing that in the definition of quasilocal mass i.e., (\ref{QLM}) 
$\eta ^{2}$\ is related to the background, the first law of thermodynamics
of the black hole simply becomes%
\begin{equation}
dM_{QL}=T_{H}dS  \label{FLT}
\end{equation}%
where the variations of $M_{QL}$\ and $S$\ are with respect to $R_{+}$ \cite%
{YZ1}.

\subsection{The spacetime structure}

\begin{figure}[tbp]
\centering\includegraphics[width=0.5\textwidth]{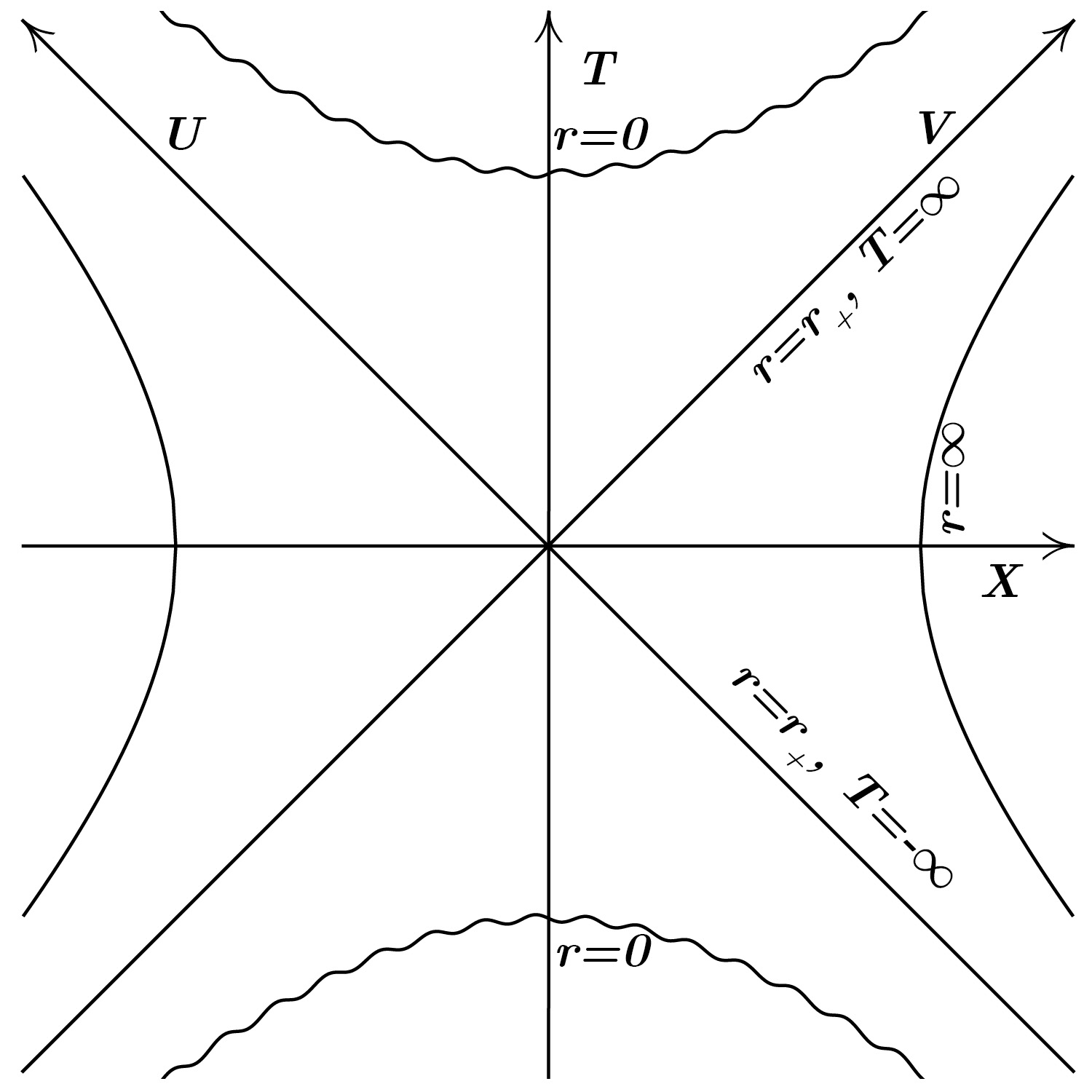}
\caption{Kruskal-Szekeres diagram of the black hole spacetime (\protect\ref%
{R2}) with ${\protect\small \protect\eta }^{2}{\protect\small =}\frac{3}{2}$%
\textbf{\ and }${\protect\small \protect\eta }^{2}{\protect\small =}2.$}
\label{F3}
\end{figure}

\begin{figure}[tbp]
\centering\includegraphics[width=0.5\textwidth]{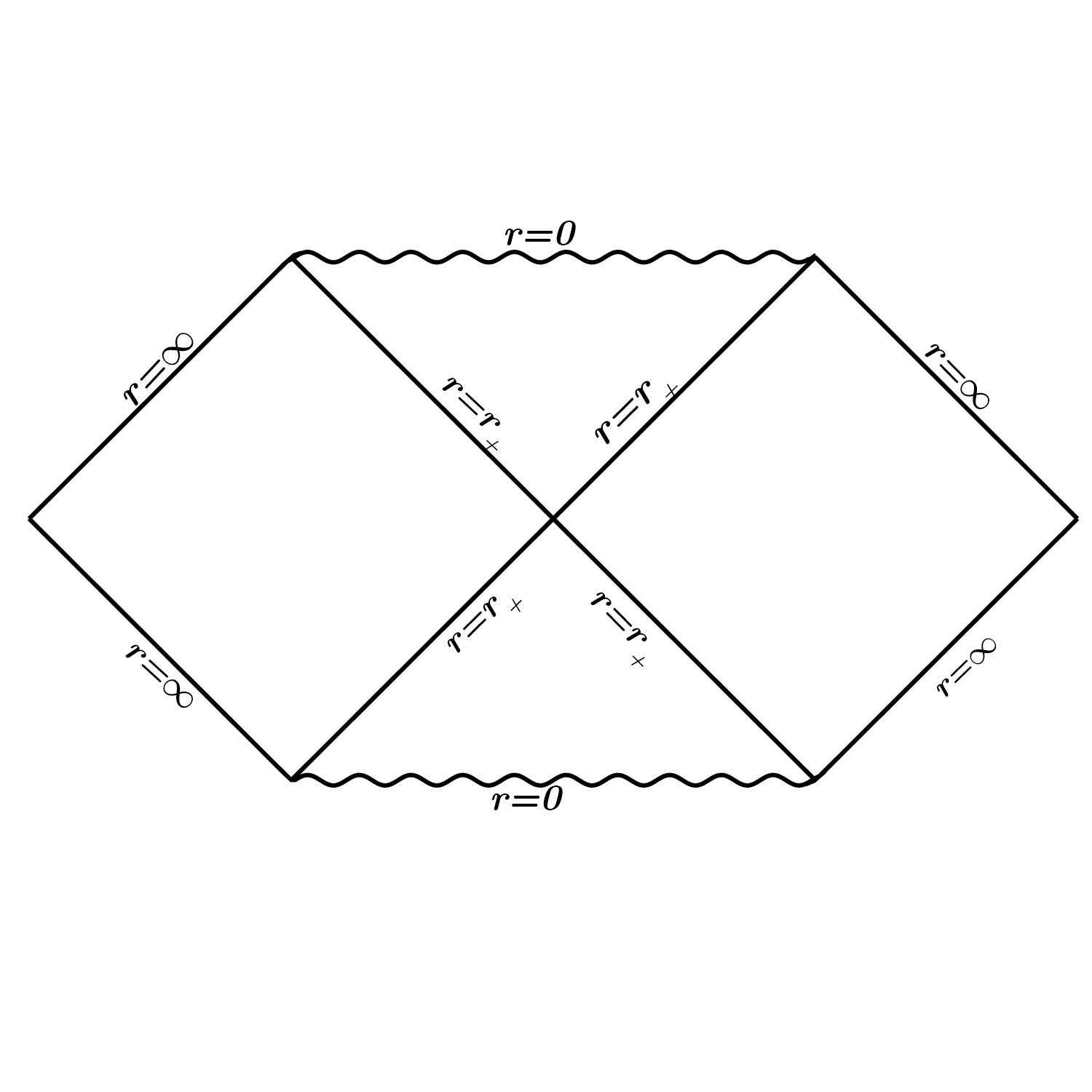}
\caption{Carter-penrose diagram of the black hole spacetime (\protect\ref{R2}%
) with ${\protect\small \protect\eta }^{2}{\protect\small =}\frac{3}{2}$%
\textbf{\ and }${\protect\small \protect\eta }^{2}{\protect\small =}2.$}
\label{F4}
\end{figure}
The spacetime described by the line element (\ref{LE}) is rather new in the
literature and deserves to be more investigated from the spacetime structure
aspect. The solution obviously is a black hole with an event horizon located
at $\rho =\rho _{+}=\left( 2M\right) ^{\frac{1}{\eta ^{2}}}$such that it may
be written as%
\begin{equation}
ds^{2}=-\left( \rho ^{\eta ^{2}}-\rho _{+}^{\eta ^{2}}\right) \rho ^{\eta
^{2}-2}d\tau ^{2}+\frac{\rho ^{\eta ^{2}}}{\rho ^{\eta ^{2}}-\rho _{+}^{\eta
^{2}}}d\rho ^{2}+\varkappa ^{2}\rho ^{2}d\Omega ^{2}.  \label{R1}
\end{equation}%
After transforming $r=\varkappa \rho $\ and redefinition of time one finds%
\begin{equation}
ds^{2}=-\left( r^{\eta ^{2}}-r_{+}^{\eta ^{2}}\right) r^{\eta ^{2}-2}dt^{2}+%
\frac{r^{\eta ^{2}}}{\varkappa ^{2}\left( r^{\eta ^{2}}-r_{+}^{\eta
^{2}}\right) }dr^{2}+r^{2}d\Omega ^{2}  \label{R2}
\end{equation}%
in which $0<r<\infty $\ and $-\infty <t<\infty $. The Kretschmann scalar of
the latter spacetime is given by%
\begin{equation}
\mathcal{K}=\frac{\omega _{0}}{r^{4}}\frac{\omega _{1}}{r^{4+\eta ^{2}}}+%
\frac{\omega _{2}}{r^{4+2\eta ^{2}}}  \label{R3}
\end{equation}%
where $\omega _{i}$\ are some constants. We recall that $1<\eta ^{2}=\frac{%
b^{2}+1}{b^{2}}$\ which implies the origin $r=0$\ is a singular point. To
find the nature of this curvature singularity we apply the so-called
conformal compactification. Hence, we obtain the conformal/tortoise radial
coordinate defined by%
\begin{equation}
r_{\ast }=\int \sqrt{-\frac{g_{rr}}{g_{tt}}}dr=\frac{1}{\chi }\int \frac{r}{%
r^{\eta ^{2}}-r_{+}^{\eta ^{2}}}dr.  \label{R4}
\end{equation}%
We observe that the conformal/tortoise radial coordinate $r_{\ast }$\
depends on $\eta ^{2}$. For technical reasons, we set $\eta ^{2}=\frac{3}{2}$
as well as $2$\ and continue our investigation accordingly. In this
configurations one finds%
\begin{equation}
r_{\ast }=\left\{ 
\begin{array}{cc}
\frac{1}{\chi }\left\{ 2\sqrt{r}+\frac{2}{3}\sqrt{r_{+}}\ln \left\vert \sqrt{%
r}-\sqrt{r_{+}}\right\vert -\frac{1}{3}\sqrt{r_{+}}\ln \left\vert r+r_{+}+%
\sqrt{r_{+}r}\right\vert -\frac{2\sqrt{3r_{+}}}{3}\arctan \left( \frac{2%
\sqrt{r}+\sqrt{r_{+}}}{\sqrt{3r_{+}}}\right) \right\} , & {\small \eta }^{2}%
{\small =}\frac{3}{2} \\ 
\frac{1}{2\chi }\ln \left\vert r^{2}-r_{+}^{2}\right\vert , & {\small \eta }%
^{2}{\small =}2%
\end{array}%
\right.   \label{R5}
\end{equation}%
where $-\infty <r_{\ast }<\infty $. Next, we define the retarded and
advanced coordinates i.e., $u=t-r_{\ast }$\ and $v=t+r_{\ast }$\ ($-\infty
<u<v<\infty $) such that (\ref{R2}) becomes%
\begin{equation}
ds^{2}=\left\{ 
\begin{array}{cc}
-r\left( 1-\left( \frac{r_{+}}{r}\right) ^{3/2}\right) dudv+r^{2}d\Omega
^{2}, & {\small \eta }^{2}{\small =}\frac{3}{2} \\ 
-r^{2}\left( 1-\left( \frac{r_{+}}{r}\right) ^{2}\right) dudv+r^{2}d\Omega
^{2}, & {\small \eta }^{2}{\small =}2%
\end{array}%
\right. .  \label{R6}
\end{equation}%
Next, we define the Kruskal-Szekeres coordinate 
\begin{equation}
U=\left\{ 
\begin{array}{cc}
\frac{4\sqrt{r_{+}}}{3\chi }\exp \left( -\frac{3\chi }{4\sqrt{r_{+}}}%
u\right) , & {\small \eta }^{2}{\small =}\frac{3}{2} \\ 
\frac{1}{\chi }\exp \left( -\chi u\right) , & {\small \eta }^{2}{\small =}2%
\end{array}%
\right.   \label{R7}
\end{equation}%
and%
\begin{equation}
V=\left\{ 
\begin{array}{cc}
\frac{4\sqrt{r_{+}}}{3\chi }\exp \left( \frac{3\chi }{4\sqrt{r_{+}}}v\right)
, & {\small \eta }^{2}{\small =}\frac{3}{2} \\ 
\frac{1}{\chi }\exp \left( \chi v\right) , & {\small \eta }^{2}{\small =}2%
\end{array}%
\right.   \label{R8}
\end{equation}%
such that ($r\geq r_{+}$)%
\begin{equation}
UV=\left\{ 
\begin{array}{cc}
\frac{16r_{+}}{9\chi ^{2}}\frac{1-\sqrt{r_{+}/r}}{\sqrt{1+r_{+}/r+\sqrt{%
r_{+}/r}}}\exp \left( 3\sqrt{\frac{r}{r_{+}}}-\sqrt{3}\arctan \left( \frac{2%
\sqrt{r/r_{+}}+1}{\sqrt{3}}\right) \right) , & {\small \eta }^{2}{\small =}%
\frac{3}{2} \\ 
\frac{\left\vert r^{2}-r_{+}^{2}\right\vert }{\chi ^{2}}, & {\small \eta }%
^{2}{\small =}2%
\end{array}%
\right. ,  \label{R9}
\end{equation}%
and $0<U<V<\infty $. The line element, hence, becomes%
\begin{equation}
ds^{2}=\left\{ 
\begin{array}{cc}
r\left( 1+r_{+}/r+\sqrt{r_{+}/r}\right) ^{3/2}e^{-3\sqrt{\frac{r}{r_{+}}}}e^{%
\sqrt{3}\arctan \left( \frac{2\sqrt{r/r_{+}}+1}{\sqrt{3}}\right)
}dUdV+r^{2}d\Omega ^{2}, & {\small \eta }^{2}{\small =}\frac{3}{2} \\ 
dUdV+r^{2}d\Omega ^{2}, & {\small \eta }^{2}{\small =}2%
\end{array}%
\right.   \label{R10}
\end{equation}%
which is regular at $r=r_{+}$\ and singular at $r=0.$\ Introducing 
\begin{equation}
U=X-T  \label{R12}
\end{equation}%
and%
\begin{equation}
V=X+T  \label{R13}
\end{equation}%
transforms the line element into%
\begin{equation}
ds^{2}=\left\{ 
\begin{array}{cc}
r\left( 1+r_{+}/r+\sqrt{r_{+}/r}\right) ^{3/2}e^{-3\sqrt{\frac{r}{r_{+}}}}e^{%
\sqrt{3}\arctan \left( \frac{2\sqrt{r/r_{+}}+1}{\sqrt{3}}\right) }\left(
-dT^{2}+dX^{2}\right) +r^{2}d\Omega ^{2}, & {\small \eta }^{2}{\small =}%
\frac{3}{2} \\ 
-dT^{2}+dX^{2}+r^{2}d\Omega ^{2}, & {\small \eta }^{2}{\small =}2%
\end{array}%
\right.   \label{R14}
\end{equation}%
such that%
\begin{equation}
X^{2}-T^{2}=UV  \label{R15}
\end{equation}%
given in Eq. (\ref{R9}). We see that the singularity at $r=0$\ corresponds
to hyperbola 
\begin{equation}
T^{2}-X^{2}=\left\{ 
\begin{array}{cc}
\left( \frac{4}{3\chi }\right) ^{2}e^{\left( -\frac{\sqrt{3}\pi }{6}\right)
}r_{+}, & {\small \eta }^{2}{\small =}\frac{3}{2} \\ 
\frac{r_{+}^{2}}{\chi ^{2}}, & {\small \eta }^{2}{\small =}2%
\end{array}%
\right.   \label{R16}
\end{equation}%
on the $TX$-plane and is valid for even $r>0$\ which implies 
\begin{equation}
T^{2}-X^{2}\leq \left\{ 
\begin{array}{cc}
\left( \frac{4}{3\chi }\right) ^{2}e^{\left( -\frac{\sqrt{3}\pi }{6}\right)
}r_{+}, & {\small \eta }^{2}{\small =}\frac{3}{2} \\ 
\frac{r_{+}^{2}}{\chi ^{2}}, & {\small \eta }^{2}{\small =}2%
\end{array}%
\right. .  \label{R17}
\end{equation}%
In Fig. \ref{F3} and \ref{F4} we plot the Kruskal-Szekeres diagram and the
maximally-extended Carter-Penrose diagram of the black hole spacetime (\ref%
{R2}) with $\eta ^{2}=\frac{3}{2}$\ and $\eta ^{2}=2$ that is also
applicable for an arbitrary $\eta ^{2}$. The nature of the singularity at
the center of the black hole is spacelike that is the same as the
Schwarzschild black hole.

\begin{figure}[tbp]
\centering\includegraphics[width=0.5\textwidth]{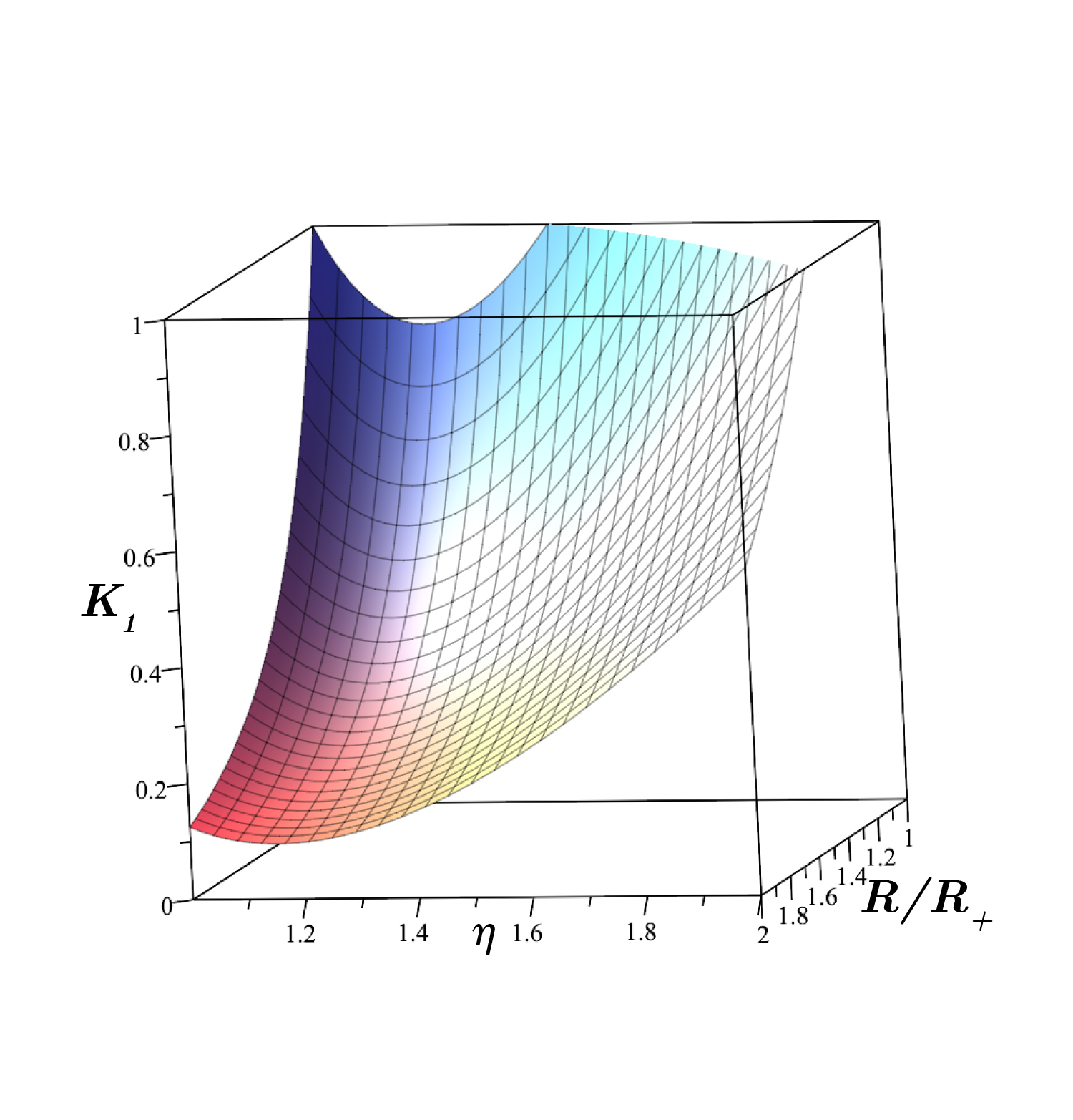}
\caption{The radial tidal force in terms of $\frac{R}{R_{+}}$ and $\protect%
\eta $. This figure implies the radial tidal force is always positive and in
terms of $\frac{R}{R_{+}}$ for a given $\protect\eta $, it is a monotonic
decreasing function that approaches zero. On the other hand for a given $%
\frac{R}{R_{+}}$, the radial tidal force first decreases in terms of $%
\protect\eta $ and then increases.}
\label{F5}
\end{figure}

\begin{figure}[tbp]
\centering\includegraphics[width=0.5\textwidth]{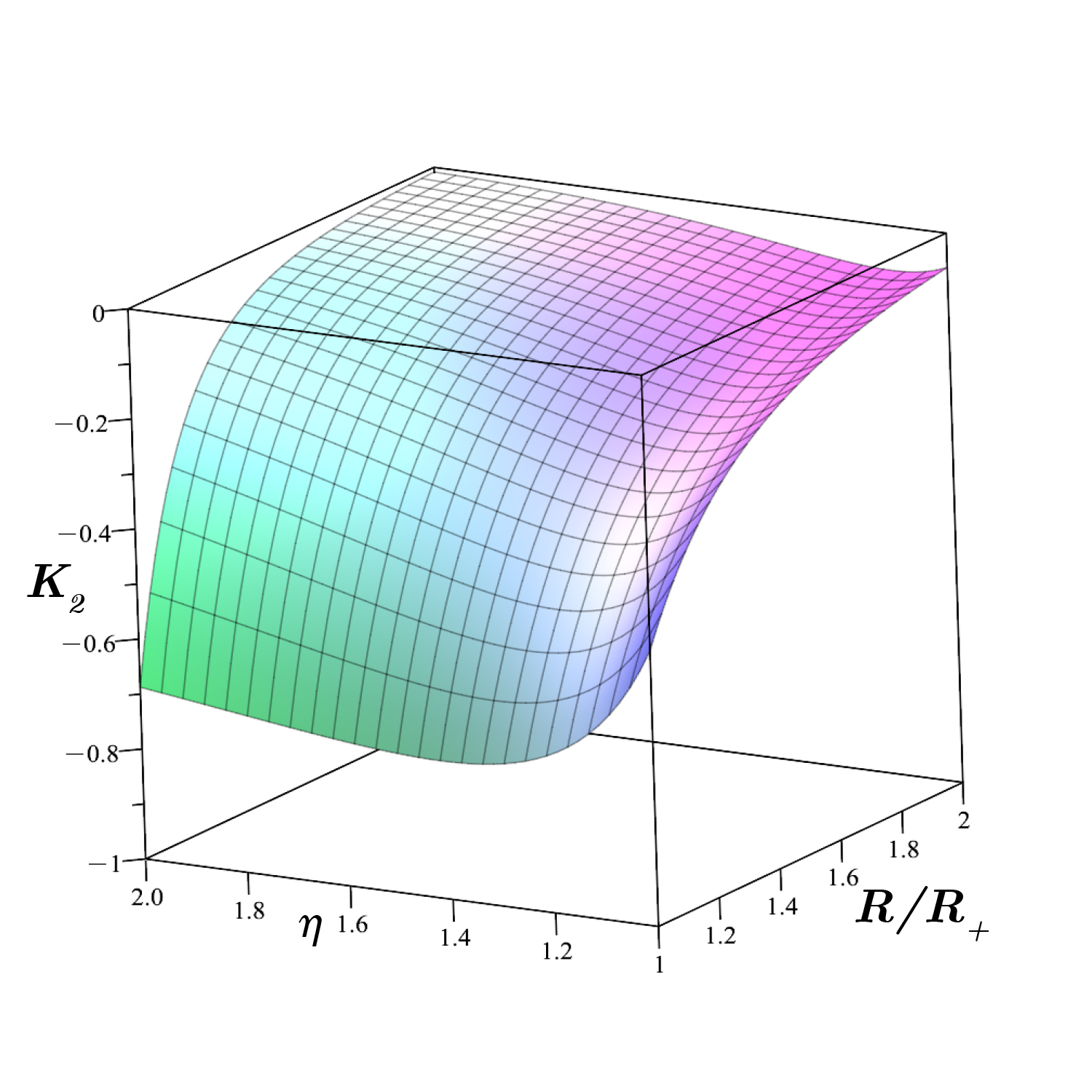}
\caption{The angular tidal force in terms of $\frac{R}{R_{+}}$ and $\protect%
\eta $. The angular tidal force is always negative and in terms of $\frac{R}{%
R_{+}}$ for a given $\protect\eta $, it is a monotonic increasing function
that approaches zero. On the other hand for a given $\frac{R}{R_{+}}$, the
angular tidal force first decreases in terms of $\protect\eta $ and then
increases.}
\label{F6}
\end{figure}

\section{Tidal forces}

In this section, we study the so-called tidal force which is an indication
of the interaction between a black hole with its surroundings. When an
extensive object falls under the gravitational attraction of a black hole,
the tidal forces exerted on the object on its geodesic cause either
stretching or compressing in different directions. For instance, when such
an object falls radially toward the Schwarzschild black hole, it is
stretched in the radial direction while compressed in the angular/transverse
directions \cite{TF1,TF2,TF3,TF4}. Unlike the Schwarzschild black hole which
is surrounded by a vacuum, in the well-known Reissner-Nordstr\"{o}m dirty
black hole for the same radially falling extensive object, the tidal forces
vanish at some certain radius and its nature changes from stretching to
compressing and vice-versa \cite{TF5}. Similar properties have also been
reported in some other black hole spacetimes where the sign-turning point
for the tidal force is either outside the event horizon or inside \cite%
{TF5,TF6,TF7,TF8}.

In \cite{TF} the tidal force tensor in the tetrad basis attached to a
radially infalling observer has been calculated for a dirty black hole with
the line element%
\begin{equation}
ds^{2}=-f\left( R\right) dt^{2}+\frac{1}{1-\frac{B\left( R\right) }{R}}%
dR^{2}+R^{2}\left( d\theta ^{2}+\sin ^{2}\theta d\varphi ^{2}\right) .
\label{R21}
\end{equation}%
In accordance with the results of \cite{TF}, the tidal force tensor is given
by%
\begin{equation}
K_{\hat{\mu}}^{\hat{\nu}}=diag\left[ 0,K_{1},K_{2},K_{3}\right]  \label{R23}
\end{equation}%
in which 
\begin{equation}
K_{1}=\frac{2B}{R^{3}}-\frac{1}{2}\left( \rho +P_{r}-2P_{t}\right)
\label{R24}
\end{equation}%
\begin{equation}
K_{2}=K_{3}=-\frac{B}{R^{3}}+\frac{1}{2}\rho -\frac{E^{2}}{2f}\left( \rho
+P_{r}\right)  \label{R25}
\end{equation}%
with $E$\ the energy per unit mass and $\rho ,$ $P_{r}$, and $P_{t}$\ given
in (\ref{EM1})-(\ref{EM3}). Note that the unit convention in \cite{TF} is $%
G=1.$\ Herein the line element is given by (\ref{M3}) such that 
\begin{equation}
f\left( R\right) =\eta ^{2}\left( 1-\left( \frac{R_{+}}{R}\right) ^{\eta
^{2}}\right) R^{2\left( \eta ^{2}-1\right) }  \label{R26}
\end{equation}%
and 
\begin{equation}
B\left( R\right) =\left( 1-\frac{1}{\eta ^{2}}\left( 1-\left( \frac{R_{+}}{R}%
\right) ^{\eta ^{2}}\right) \right) \frac{R}{2}.  \label{R27}
\end{equation}

In Fig. \ref{F5} and \ref{F6}, we plot $K_{1}$\ and $K_{2}$\ in terms of $%
\frac{R}{R_{+}}$\ ($\frac{R}{R_{+}}>1$) and $\eta $\ ($1<\eta $) and $E=1.$\
We observe that similar to the tidal force in a Schwarzschild black hole,
the radial tidal force is tensile and the transverse is compressive. With
increasing the value of $\eta $\ first both forces decrease and then
increase and both approach zero at infinity.

\section{Conclusion}

In the framework of Einstein's gravity coupled to SR-NED as well as a
Dilaton field, we managed to solve the field equations exactly and obtain a
black hole solution characterized by two parameters namely, $R_{+}$ and $b$.
While the latter is a theory constant representing the Dilaton the former is
an integration. This non-asymptotically flat black hole is singular at its
center where the electric charge is placed. The electric field is radial but
uniform in the sense that the electromagnetic invariant is a constant. Since
the ADM mass is not defined for non-asymptotically flat black holes, we
applied BY formalism to obtain the QL conserved mass expressed as $M_{QL}$\
such that in the Schwarzschild limit ($\eta ^{2}=1$), it coincides with the
ADM mass of the Schwarzschild black hole. Furthermore, we studied the null
geodesic on the equatorial plane and showed that the fate of photons depends
on the ratio $\frac{\mathcal{E}^{2}}{\ell ^{2}}$ where $E$ and $\ell $ are
the conserved energy and angular momentum. Moreover, we investigated the
thermal stability of the black hole and observed that with $0<b^{2}<1$ ($%
2<\eta ^{2}$) the black hole is thermally stable in the sense that both the
Hawking temperature and the heat capacity are positive. As it was stated in 
\cite{BR1}, black holes are not forming in the empty space and rather are
surrounded by matter fields that are either falling into the black holes or
are in equilibrium with it. The black hole presented in this paper is a
typical example of such a black hole called a "dirty" black hole supported
by normal/regular matter. The mathematical structure of the spacetime is
kind of modified Schwarzschild black hole because with $\eta ^{2}=1$\ it
coincides with the Schwarzschild black hole. Therefore, one may call this
solution the natural dirty Schwarzschild black hole. Concerning what we have
done in this study we may consider the following to be the novelty of our
paper: i) We filled the gap of the PM-NED model which has so far been
considered with $p\neq \frac{1}{2}$. ii) The black hole in the context of
our study has been found to be a new generalization of the Schwarzschild
black hole in a practically simple form. We have shown that its casual
structure is similar to the Schwarzschild black hole as well. iii) The
dirty-Schwarzschild - this is what we named our solution - the black hole is
thermally stable for $2<\eta ^{2}$. iv) We calculated the tidal force which
clearly is in agreement with the Schwarzschild black hole although its
minimum value takes place in $\eta ^{2}\neq 1,$ recalling that $\eta ^{2}=1$%
\ is the Schwarzschild limit.

\section*{Acknowledgement}

The author would like to thank the anonymous reviewers for their
constructive comments which improved the manuscript significantly.

\end{document}